\documentclass[prl,twocolumn,superscriptaddress]{revtex4-2}
\usepackage{amsmath,amssymb,mathrsfs}
\usepackage{psfrag}
\usepackage[normalem]{ulem}
\usepackage{graphicx}
\usepackage{graphics}
\usepackage{epsfig}
\usepackage{bm}
\usepackage{color}
\usepackage{verbatim,color}
\usepackage{physics}
\usepackage{hyperref}
\usepackage[normalem]{ulem}
\usepackage{cancel}

\newcommand{\braketmatrix}[3]{\left \langle #1 \middle| #2 \middle| #3 \right \rangle}
\usepackage[acronym]{glossaries}
\newcommand{\beq}{\begin{equation}}
\newcommand{\eeq}{\end{equation}}
\newcommand{\beqa}{\begin{eqnarray}}
\newcommand{\eeqa}{\end{eqnarray}}
\newcommand{\ba}{\begin{aligned}[b]}
\newcommand{\ea}{\end{aligned}}

\newcommand{\cred}{\color{black}}

\begin{document}

\title{\cred{Macroscopic quantum self-trapping in bosonic Josephson junctions:  \\ an exact quantum treatment}}
\author{A. Bardin}
\affiliation{Dipartimento di Fisica e Astronomia "Galileo Galilei", 
Universit\`a di Padova, Via Marzolo 8, 35131 Padova, Italy}
\affiliation{Istituto Nazionale di Fisica Nucleare (INFN), Sezione di Padova, via Marzolo 8, 35131 Padova, Italy
}
\author{A. Minguzzi}
\affiliation{
Université Grenoble Alpes,  CNRS, LPMMC,  38000 Grenoble, France}
\author{L. Salasnich}
\affiliation{Dipartimento di Fisica e Astronomia "Galileo Galilei", 
Universit\`a di Padova, Via Marzolo 8, 35131 Padova, Italy}
\affiliation{Istituto Nazionale di Fisica Nucleare (INFN), Sezione di Padova, via Marzolo 8, 35131 Padova, Italy
}
\affiliation{Istituto Nazionale di Ottica (INO) del Consiglio Nazionale delle Ricerche (CNR), via Nello Carrara 1, 50019 Sesto Fiorentino, Italy}

\begin{abstract}
{\cred We investigate the fully quantum evolution of the population imbalance in a perfectly symmetric Bose-Josephson junction modeled by a two-mode Bose-Hubbard Hamiltonian, focusing on the validity of macroscopic quantum self-trapping beyond the mean-field theory. We show that for any finite number of particles the exact quantum dynamics leads to the breakdown of macroscopic quantum self-trapping after a finite time, regardless of the initial state. Using the symmetries of the Bose-Hubbard Hamiltonian, we provide a mathematical demonstration of this result and analyze the spectral properties governing the dynamics. We identify a branching behavior in the eigenvalues differences and a nontrivial structure of the population-imbalance amplitudes. These features allow us to distinguish two clearly different dynamical regimes and to elucidate the mechanism leading to the emergence of a quasi-MQST regime for large particle numbers. These findings bridge the gap between mean-field predictions and exact quantum dynamics and provide insight into the emergence of classical nonlinear behavior from finite quantum many-body systems.

} 

\end{abstract}

\maketitle

The Josephson effect \cite{josephson1962} is a striking example of a macroscopic quantum phenomenon,  evidenced by coherent oscillations among two  superconductors or superfluids separated by a tunneling barrier.
Superconducting Josephson junctions play a significant role in modern technologies \cite{Wolf2017}: for instance, superconducting quantum interference devices (SQUIDs) serve as highly sensitive magnetometers, and Josephson junctions are used for the implementation of qubits in quantum computing \cite{Ladd2010,Buluta2011}.
The first realization of Bose-Einstein condensates with ultracold dilute alkali-metal atoms \cite{Cornell1995,Ketterle1995}  paved the way for an alternative type of Josephson junction based on ultracold bosonic atomic gases, named  Bose-Josephon junctions.  First theoretically proposed \cite{Smerzi1997} and later experimentally realized \cite{Albiez2005,Shin2005,Levy2005}, these systems consist of two weakly linked Bose–Einstein condensates trapped in a double‐well potential.

At difference from superconducting junctions, two distinct dynamical regimes can be accessed in Bose-Josephson junctions.
First, for small initial population imbalance and large tunnel amplitude,  a Josephson oscillation regime occurs, characterized by sinusoidal oscillations of relative phase and population imbalance around a zero mean value, with frequency corresponding to the Josephson frequency.  A second more elusive regime occurs for large initial population  imbalance, denoted macroscopic quantum self-trapping (MQST) \cite{Smerzi1997} and
characterized by small-amplitude oscillations of the population imbalance around a non-zero mean value, while the relative phase increases over time.
These two regimes are obtained within a two-mode mean-field theory, which holds in the limit of large particle numbers $N$ and weak interactions \cite{Smerzi1997, Smerzi1999}. The mean-field approximation, however,  neglects many-body effects that crucially emerge in the quantum regime, at larger interactions and in finite-size systems.
 The very same occurrence of MQST beyond mean field is so far not fully elucidated. Interactions with non-condensed particles have been shown to lead to decoherence and dephasing, undermining the stability of self-trapped dynamics \cite{PhysRevA.58.R50,PRL118.230403}. The atomic coherent state approach provides improved finite-size corrections to mean-field dynamics but fails to reliably predict the onset of macroscopic quantum self-trapping, which remains accurately captured only in the large-$N$ limit \cite{Wimberger2021}. Moreover, systematic inclusion of Gaussian quantum corrections shows that quantum fluctuations modify both the Josephson frequency and the MQST critical interaction strength, with these effects strongly dependent on dimensionality \cite{Bardin2024}. 
 
In this Letter, we address the question of the existence of MQST beyond mean field through a fully quantum analysis of the strongly out-of-equilibrium dynamical evolution of a Bose-Josephson junction. Our work is organized into two parts. First
we provide a no-go theorem, i.e. we prove mathematically that MQST does not occur for any finite number of particles in the system, {\cred{ independently of the initial state}}. Then, we conciliate this mathematical statement with 
the mean-field limit, by identifying 
 a phase transition in the energy-level spacings of the spectrum  of the quantum Hamiltonian, as obtained 
 by exact diagonalization, as well as a change of behaviour in the coefficients of the dynamical evolution of the population imbalance. This shows that there are two very distinct dynamical regimes in the quantum model, which we link to the MQST and the Josephson regimes of the mean-field solution.

{\it Model}— We describe the bosonic Josephson junction within the two-mode approximation by the two-site Bose-Hubbard Hamiltonian \cite{Milburn1997,Ferrini2008}, {\cred{which is exactly solvable by Bethe ansatz methods} \cite{Links2006}}:
\begin{equation}
    \begin{split}
        \hat{H}=&E_1^0\hat{a}_1^\dagger\hat{a}_1+E_2^0\hat{a}_2^\dagger\hat{a}_2+\frac{U_1}{2}\hat{a}_1^\dagger\hat{a}_1^\dagger\hat{a}_1\hat{a}_1+\\
        &+\frac{U_2}{2}\hat{a}_2^\dagger\hat{a}_2^\dagger\hat{a}_2\hat{a}_2-\frac{J}{2}\left(\hat{a}_2^\dagger\hat{a}_1+\hat{a}_1^\dagger\hat{a}_2\right)
    \end{split}
    \label{HBH}
\end{equation}
Here, $\hat{a}_j, \hat{a}_j^\dagger$ with $j = 1,2$ are bosonic creation (annihilation) operators obeying the canonical commutation relations $[\hat{a}_j, \hat{a}_\ell^\dagger] = \delta_{j\ell}$, $E_j^0$ are the single-particle energies of the two wells in the double-well potential, $U_j>0$ denotes on-site repulsive interactions, and $J/2$ is the tunneling amplitude between the wells.

To link with Josephson effect, it is useful to express
 the field operators in number-phase representation 
 as $ \hat{a}_j=\sqrt{\hat{N}_j+1}\hat{e}^{i\phi_j}$ and $\hat{a}_j^\dagger=\hat{e}^{-i\phi_j}\sqrt{\hat{N}_j+1}$ \cite{Susskind1964}, 
where $\hat{N}_j=\hat{a}_j^\dagger\hat{a}_j$ is the number operator for the mode $j$, and $\hat{e}^{-i\phi_j}$ is the operator corresponding to the exponential of the phase,  satisfying the commutation relation $[\hat{N}_j,\hat{e}^{i\phi_j}]=-\hat{e}^{i\phi_j}$.
In the following,  we will work at fixed total number $N$ of bosons, where $N$ is the eigenvalue of ${\hat N}={\hat N}_1+{\hat N}_2$. By defining the relative-phase operator $\hat{e}^{i\phi}=\hat{e}^{i\phi_2}\hat{e}^{-i\phi_1}$ and  the relative number  operator $\hat{n}=\left(\hat{N}_1-\hat{N}_2\right){/2}$, the Hamiltonian (\ref{HBH}) can be recast as \cite{Ferrini2008}
\begin{equation}
\begin{split}
    \hat{H}=&U\left(\hat{n}-n_0\right)^2+\\
    &-\frac{J}{4}\left(\sqrt{N+\hat{n}+2}\sqrt{N-\hat{n}}\hat{e}^{-i\phi}\right.\\
    &\left.+\sqrt{N-\hat{n}+2}\sqrt{N+\hat{n}}\hat{e}^{i\phi}\right).
\end{split}
\label{H_BHH}
\end{equation}
Here $n_0=-\Delta\tilde{E}/2U$ is related to the effective well asymmetry  $\Delta\tilde{E}=E_1^0-E_2^0+(U_1-U_2)(N-1)/2$, and $U=(U_1+U_2)/2$ is the average interaction strength.
Working at fixed $N$, the eigenstates $\ket{n}$  
of ${\hat n}$ are such that $n\in [-N/2,N/2]$.
Since the relative-phase operator can be represented as $\hat{e}^{i\phi}=\sum_{n=-N/2+1}^{N/2}\ket{n-1}\bra{n}$ \cite{Susskind1964}, the Hamiltonian can be represented as a tridiagonal matrix with the following  matrix elements, expressed
in units of $J$:
\begin{equation}
\begin{split}
    \mel{n}{\hat{H}}{n}&=\frac{\Lambda}{N}(n-n_0)^2\\
    \mel{n\pm1}{\hat{H}}{n}&=-\frac{N}{2}\sqrt{\frac{1}{2}\pm\frac{n\pm1}{N}}\sqrt{\frac{1}{2}\mp \frac{n}{N}}
    \label{H}
\end{split}
\end{equation}
where $\Lambda=UN/J$ 
is the dimensionless coupling  strength.
In the following, we will consider the symmetric double-well potential, where $n_0=0$.

{\it No-go theorem}— We first demonstrate that, {\cred{in the completely symmetric double-well potential case, i.e., $n_0=0$ case}}, the macroscopic quantum self-trapping  cannot occur in the fully quantum evolution for any finite number of particles, {\cred{independently of the choice of the initial state}}. 

To this end, we evaluate the time-dependent expectation value of the population imbalance $\langle \hat{z}(t) \rangle$, where  $\hat{z}= { 2} \, \hat n /N$,
starting from {\cred{a generic}} state $\ket{\alpha_0}$. 

We adopt the Schrödinger picture and express the time-evolved state
under the fully quantum evolution as  $\ket{\alpha(t)} = e^{-i\hat{H}t/\hbar} \ket{\alpha_0}$. 
It is then advantageous to decompose the initial state in terms of the energy eigenbasis of the Hamiltonian, {\cred{namely $\ket{k}$ with $0\leq k\leq N$ with eigenstate $E_k$. In this way, the expectation value of the  population imbalance is given by
\begin{equation}
\begin{split}
    \langle \hat{z}(t)\rangle=&\frac{1}{2}\sum_{k=0}^NA_{kk}C_{kk}+\sum_{k=1}^N\sum_{k'=0}^{k-1}A_{kk'}C_{kk'}\cos{\left(\frac{E_{kk'}}{\hbar}t-\theta_{kk'}\right)}
\end{split}
\end{equation}
where $A_{kk'}\equiv\frac{4}{N}\braketmatrix{k}{\hat{n}}{k'}$ and  $E_{kk'}\equiv E_k-E_{k'}$ depend on the Hamiltonian eigenvectors and eigenvalues while in $C_{kk'}=|c_k||c_{k'}|$ and $\theta_{kk'}=\arg(c^*_kc_{k'})$ it is encoded the initial state dependence, since $c_k\equiv\braket{\alpha_0}{k}$. The macroscopic quantum self-trapping regime is  characterized by an oscillation of the population imbalance around a non-equilibrium value. Let us first consider the completely symmetric case, i.e. $n_0=0$. 
}}
The Hamiltonian $\hat{H}$ is not only symmetric, i.e. $
    \mel{n}{\hat{H}}{n \mp 1} = \mel{n \mp 1}{\hat{H}}{n}$, 
but also centrosymmetric since $\mel{n \mp 1}{\hat{H}}{n} = \mel{-n \pm 1}{\hat{H}}{-n}$ and $  \mel{n}{\hat{H}}{n} = \mel{-n}{\hat{H}}{-n}$.
As shown in Refs.~\cite{Andrew1973,Cantoni1976}, symmetric centrosymmetric $(N+1)\times(N+1)$ matrices with even $N$ have eigenvectors that can be ordered as alternating symmetric and antisymmetric, beginning with the symmetric one. Details are given in Supplemental Material~\cite{SupMat}. 
Thus, the eigenbasis $\{\ket{k}\}$ splits into two orthogonal families: skew-symmetric states $\ket{i}$ with $\hat{H} \ket{i} = \lambda_i \ket{i}$ and $v_{i,n} = -v_{i,-n}$, and symmetric states $\ket{j}$ with $\hat{H} \ket{j} = \ell_j \ket{j}$ and $w_{j,n} = w_{j,-n}$.

Using this decomposition, the expectation value of the population imbalance becomes
{\cred{
\begin{equation}
    \langle \hat{z}(t) \rangle = \sum_{i=1}^{N/2} \sum_{j=0}^{N/2} A_{ij} C_{ij} \cos\left(\frac{E_{ij}}{\hbar} t-\theta_{ij}\right),
    \label{eq:zt}
\end{equation}
}}

Since $\langle \hat{z}(t) \rangle$ is a sum of cosines, persistent non-zero imbalance, i.e., full MQST, would require that there exists a set of indices $(i,j)$ such that {\cred{$E_{ij}=0$}} with non-zero weights {\cred{$A_{ij}C_{ij}$}}. However,  for a real symmetric tridiagonal  matrix, if all its off-diagonal elements are non-zero, which is the case of the Hamiltonian  (\ref{H})  for any finite number of particles, all the eigenvalues are non-degenerate \cite{Parlett1997}. Hence, $z(t)$ is doomed to vanish in a finite time, i.e., the function $z(t)$ has an oscillatory behavior and crosses the $z=0$ axis. This shows that MQST cannot occur in the quantum model for any finite $N$.
We finally notice that, upon normalizing the Hamiltonian   (\ref{H}) by $N/2$, the off-diagonal entries are bounded between $-\sqrt{1 + 2/N}/2$ and  $-1/\sqrt{N}$. 
In the limit $N \to \infty$, the last off-diagonal term vanishes, allowing for degeneracies that may support MQST. We show {\cred{in the next section}}  how the quantum analog of MQST emerges from the large-$N$ limit  of the quantum model under specific interaction conditions.
{\cred{In the asymmetric case, i.e. $n_0\neq0$, a generalization of the theorem is not granted, as $A_{kk}$ terms are not constant as $k$ change and therefore there are more than one equilibrium configuration.}}

{\it Branching transition}—
Let us reconsider Eq. (\ref{eq:zt}) for the population imbalance expectation value and analyze how energy level differences and matrix elements  scale  with particle number and interaction strength.
It is convenient to express the indices $(i,j)$ in Eq.~(\ref{eq:zt}) in terms of their  sum, $\sigma \equiv i + j$,   and  difference, $\delta \equiv j - i$. 
{\cred{Clearly, $1\leq \sigma \leq N$ while $-N/2\leq \delta \leq (N/2)-1$.}} 
In these new variables, the population imbalance reads
\begin{equation}
     \langle\hat{z}(t)\rangle=\sum_{\delta=-N/2}^{N/2-1}\sum_{\substack{\sigma =\max( 2+\delta,-\delta) \\ \sigma \equiv \delta \pmod{2}}}^{N-|\delta|}A_{\delta,\sigma}C_{\delta,\sigma}\cos{(E_{\delta,\sigma}t/\hbar)} .
     \label{eq:zt-delta-sigma}
\end{equation}

\begin{figure}
    \centering
    \includegraphics[width=1.\linewidth]{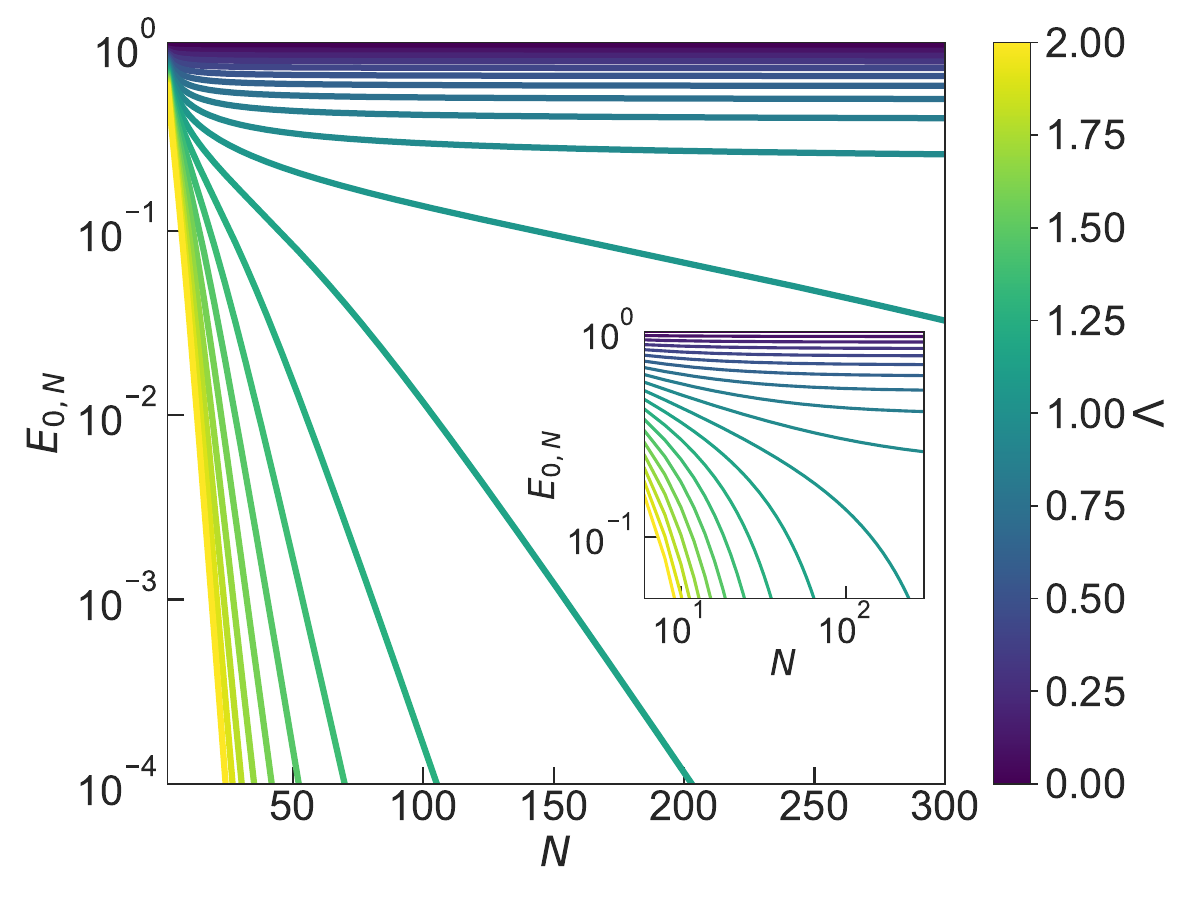}
    \caption{Eigenvalues difference $E_{0,N}\equiv\lambda_{N/2}-\ell_{N/2}$ as a function of particle number $N$ for different values of the dimensionless interaction strength $\Lambda=UN/J\in[0,2]$ as indicated on the color bar, in semi-logarithmic scale, showing exponential decrease at large $\Lambda$. The inset shows the same data in a log-log scale to highlight the power-law decay at small $\Lambda$ values.}
    \label{fig:E_0N}
\end{figure}

Let us first focus on the energy differences $E_{\delta,\sigma}$. We have demonstrated above that full MQST is unattainable at any finite $N$  due to the absence of degenerate eigenvalues. However, examining the energy difference for $\delta = 0$ and $\sigma = N$, corresponding to the family of eigenvalues which have the smallest differences,  we observe that its dependence on the particle number $N$ exhibits two distinct regimes -- see Fig.~\ref{fig:E_0N}.  For small values of the interaction strength, i.e. for  $\Lambda \lesssim 1$, the eigenvalue differences decay as a power-law, whereas for larger interaction strengths $\Lambda \gtrsim 1$, it decays  exponentially. Thus, a critical interaction strength $\Lambda \sim 1$ separates these two behaviors.
\begin{figure}
    \centering
    \includegraphics[width=1.\linewidth]{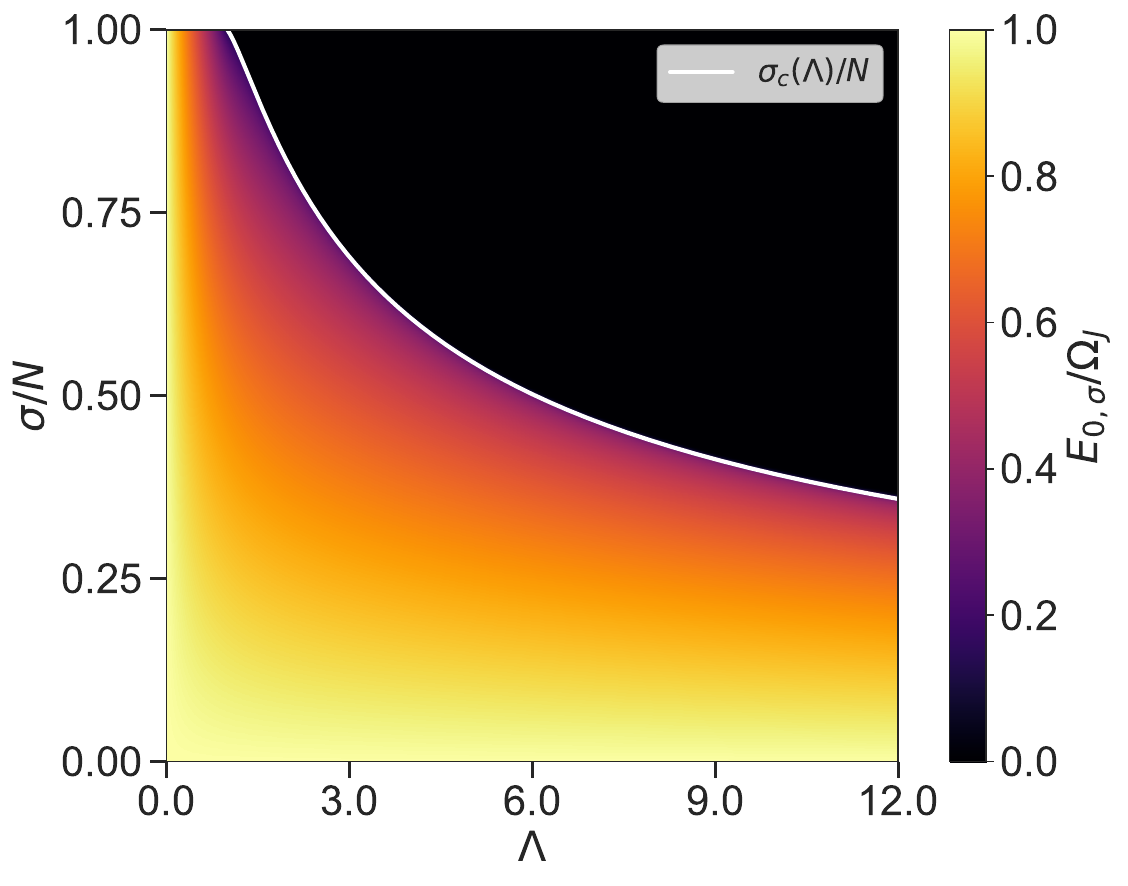}
    \caption{Color map of the eigenvalue difference $E_{0,\sigma}$ defined in the text
    in units of the  mean-field Josephson frequency
$\Omega_J =J\sqrt{1+\Lambda}$,  as a function of the
interaction strength $\Lambda$ and the normalized sum index $\sigma/N$ for $N = 500$ particles. The solid gray curve corresponds to the branching transition line,  and is given in Eq~(\ref{sigmacrit}).}
    \label{fig:E(L,s)}
\end{figure}
As a result, in a system with large particle numbers $N\gg 1$ the energy difference $E_{0,N}$ remains finite for small $\Lambda$, while it becomes vanishingly small at larger $\Lambda$. This is illustrated in Fig.~\ref{fig:E(L,s)}, where for large particle numbers the empirical relation
\begin{equation}
\frac{\sigma_c(\Lambda)}{N}\simeq\frac{\Lambda-1}{\varpi\Lambda\sqrt{2\Lambda-1}}+\frac{1}{\sqrt{\Lambda}}
    \label{sigmacrit}
\end{equation}
with $\varpi \simeq 2.622$ the lemniscate constant, well defines a critical sum index $\sigma_c(\Lambda)$ that separates regions with finite {\cred{($\sigma < \sigma_c(\Lambda)$)}}
and infinitesimal 
{\cred{($\sigma > \sigma_c(\Lambda)$)}} 
values of $E_{0,\sigma}$. 
 This behavior holds for general values of the sum index $\sigma$; however, for smaller values of $\sigma$, the critical interaction strength increases.
The fact that some energy differences become infinitesimal suggests the existence of a quasi-MQST, characterized by oscillations of $\langle \hat{z}(t) \rangle$ around a nonzero value that persist for a considerably long -- though finite --  time; this resembles the fact that {\cred{for a symmetric double-well
potential, a finite-energy particle initially localized in one well will eventually tunnel
to the other well in finite time. }} In our system, this occurs when, at a given $\Lambda$, the terms with $\sigma > \sigma_c(\Lambda)$ dominate the sum in Eq.(\ref{eq:zt-delta-sigma}) via their associated coefficients $A_{0,\sigma} C_{0,\sigma}$, whose behavior we analyze.
  As shown in Fig.~\ref{fig:AC}, at fixed initial population imbalance $z_0$, the product is peaked around a value $\sigma^*$, which increases with interaction strength $\Lambda$. The function $\sigma^*(\Lambda)$ is not continuous; rather, it exhibits a jump at a critical value $\bar \Lambda$, which depends on the initial imbalance $z_0$. Regardless of $z_0$, for $\Lambda < \bar \Lambda$, the dominant contributions to the sum come from terms with $\sigma^* < \sigma_c$, i.e., those with finite $E_{0,\sigma}$. Conversely, for $\Lambda > \bar \Lambda$, the dominant terms correspond to $\sigma^* > \sigma_c$, i.e., those with infinitesimal $E_{0,\sigma}$  (solid lines in Fig.\ref{fig:AC}). Therefore, $\bar \Lambda$ can be interpreted as a quantum generalization of the mean-field critical value $\Lambda_{c,\mathrm{MF}}$. $\Lambda_1$ and $\Lambda_2$ represent other ways of identifying the critical interaction strength from the quantum model (see Supplemental Material for details \cite{SupMat}).
Notice that $\bar \Lambda$ remains finite for $N\rightarrow \infty$ \cite{SupMat}.
The results shown here are obtained for the sake of generality by taking as initial state the ground state of the Hamiltonian in Eq.$(\ref{H})$, prepared with interaction strength $\Lambda_0$ and a non-zero well asymmetry $\tilde{z}_0=2n_0/N \neq 0$; however, very similar results are found starting from atomic coherent states (see  \cite{SupMat}).

\begin{figure}
    \centering
    \includegraphics[width=1.\linewidth]{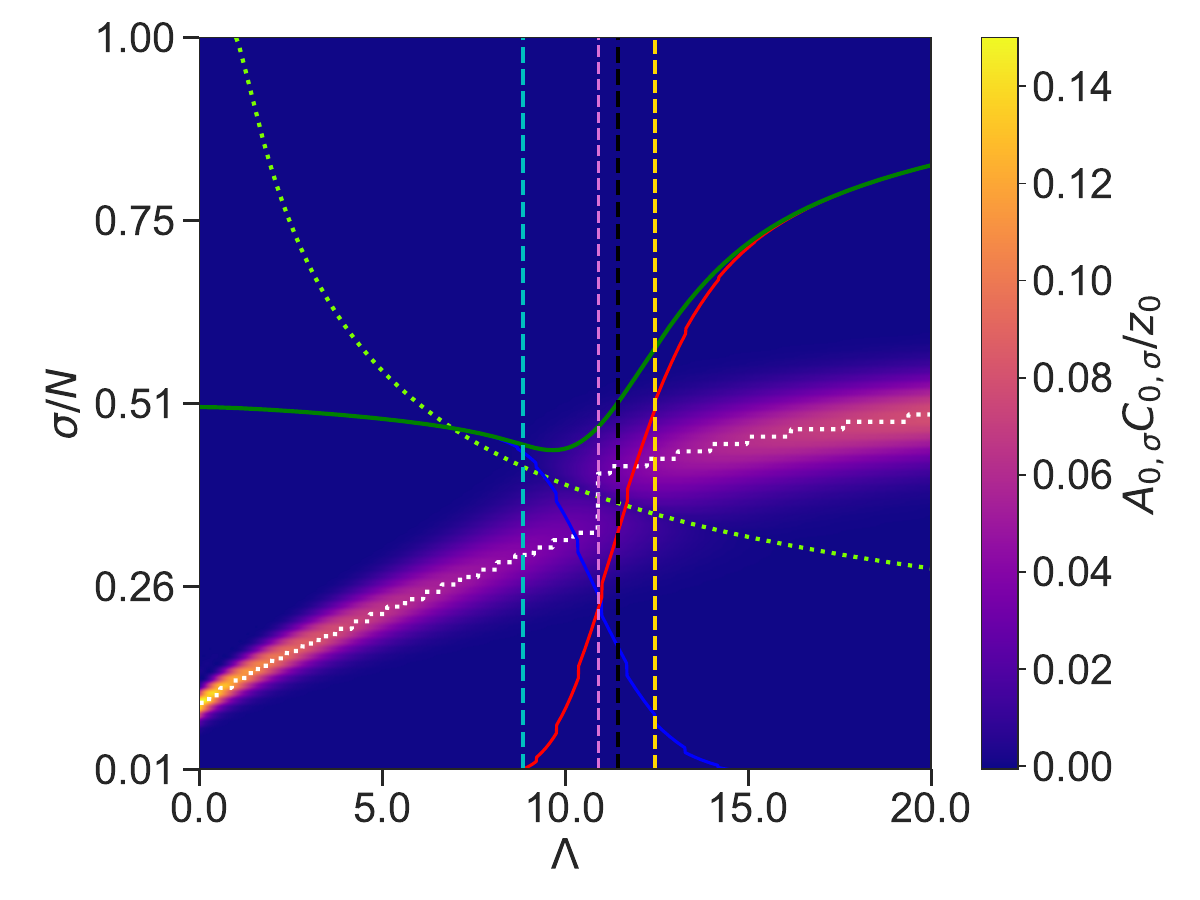}
    \caption{
Color map of the product of the coefficients $A_{0,\sigma}C_{0,\sigma}$ normalized by the initial population imbalance $z_0$, as a function of the dimensionless interaction strength $\Lambda$ and the normalized sum index $\sigma/N$ for a system with $N=200$ particles. The white dotted line indicates the value $\sigma^*/N$ corresponding to its maximum for each value of $\Lambda$. The lime dotted line shows the critical sum index \eqref{sigmacrit}. 
The three solid lines correspond to the sum of $A_{0,\sigma}C_{0,\sigma}/z_0$ over all $\sigma$ (blue), over $\sigma > \sigma_c$ (red), and over $\sigma < \sigma_c$ (green). 
The vertical dashed lines indicate   critical interaction strengths for MQST in mean field ($\Lambda_{c,\mathrm{MF}}$, black) and for quasi-MQST  ($\bar \Lambda$, magenta; $\Lambda_1$, cyan;  $\Lambda_2$ yellow).  The initial state is prepared with $\Lambda_0=10$ and $\tilde{z}_0=0.6$. 
}
    \label{fig:AC}
\end{figure}

\begin{figure}
    \centering
    \includegraphics[width=1.\linewidth]{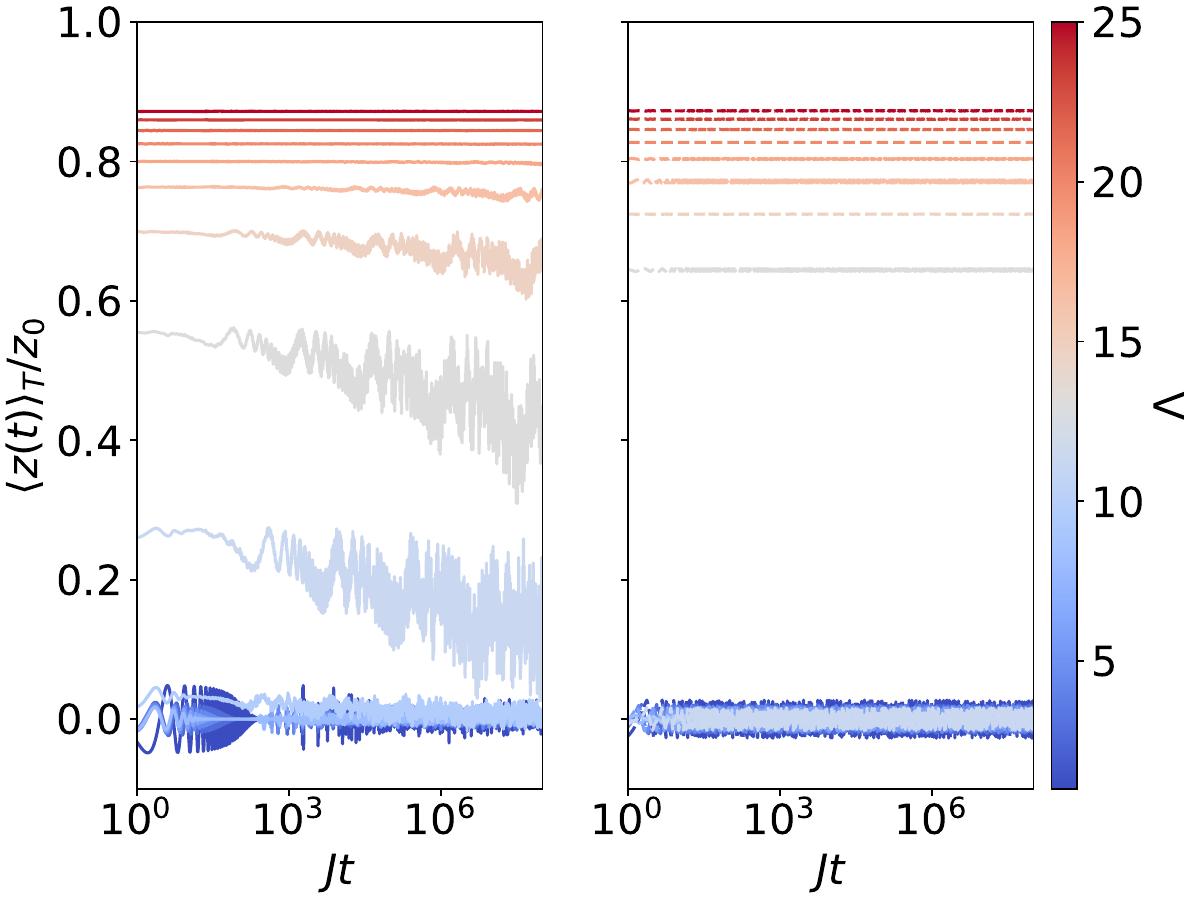}
    
    \caption{Fully quantum  (left panel) and mean-field (right panel) time evolution of the time-averaged population imbalance   $\langle z(t)\rangle_T / z_0$ (dimensionless) as a function of time, in unites of $J^{-1}$, for a system of $N=200$ particles and initial imbalance $z_0 = 0.57$,  at various values of the interaction strength $\Lambda$  as indicated in the color-bar label.  The running averaging time  window is set to $JT = 25$. The initial state of the quantum problem is the ground state of the Hamiltonian with parameters $\Lambda_0 = 20$ and $\tilde{z}_0 = 0.6$, and for the mean-field we have taken $z(0)=z_0$ and $\phi(0)=0$. }
    \label{fig:z_t}
\end{figure}

The above analysis of both eigenvalues and coefficients allows us to identify two very different dynamical regimes at large $N$ depending on interaction strength $\Lambda$.  As illustrated in Fig.~\ref{fig:z_t} (left panel),  the two regimes clearly emerge in the fully quantum dynamical evolution of population imbalance, upon filtering very rapid time oscillations by 
 $   \langle z(t) \rangle_T = (1/T) \int_t^{t+T} \langle z(\tau) \rangle \, d\tau.$
 At low interaction strength, for $\Lambda < \bar \Lambda$  we identify the analog of the Josephson regime, while for $\Lambda > \bar \Lambda$ we observe a quasi-MQST regime with non-zero values of population imbalance persisting till very long times. These two regimes are close to the mean-field ones, obtained using \cite{Smerzi1999} (reported for comparison in the right panel). However, quantum and mean-field predictions strongly differ at the crossover point, i.e. for  values  $\Lambda \sim  \bar \Lambda$.

{\it Conclusion}—
We have analyzed the fully quantum evolution of the population imbalance in a two-mode Bose-Josephson. We have first shown mathematically that no macroscopic quantum self-trapping can occur for any finite particle number, since the population imbalance is an oscillating function of time, as follows from a property of non-degeneracy of the eigenvalues of the quantum Hamiltonian. We have then tracked the emergence of MQST in the mean-field limit as being associated with a branching transition at very large $N$  of the eigenvalue differences, as well as a jump and change of nature of the overlaps with the initial state at a critical interaction strength which emerges in the quantum problem. Our analysis is confirmed by exact diagonalization simulations of the full quantum dynamics, highlighting for large $N$ a transition between a Josephson regime and a quasi-MQST one at increasing interactions. 

Our work proposes a new approach to explore the classical to quantum transition. It may be generalized to other quantum systems, providing a way to pinpoint the truly quantum nature of their dynamical evolution. State-of-art experiments in Bose-Josephson junctions with ultracold atoms may access the crossover regime from Josephson to quasi-MQST  by tuning the dimensionless coupling strength $\Lambda$, where we predict strong deviations from the mean-field predictions. This would provide evidence of the branching transition. 

{\it Acknowledgments}—
This work is partially supported by the ‘IniziativaSpecifica
Quantum’ of INFN, by the European Union-NextGenerationEU within the National Center for HPC, Big
Data and Quantum Computing (Project No. CN00000013, CN1 Spoke 10: ‘Quantum Computing’), by the
EU Project PASQuanS 2 ‘Programmable Atomic Large-Scale Quantum Simulation’, and the National Grant
of the Italian Ministry of University and Research for the PRIN 2022 project ‘Quantum Atomic Mixtures:
Droplets, Topological Structures, and Vortices’.

\clearpage
\onecolumngrid

\setcounter{equation}{0}
\setcounter{section}{0}
\setcounter{figure}{0}
\renewcommand{\theequation}{S\arabic{equation}}
\renewcommand{\thesection}{S\arabic{section}}
\renewcommand{\thefigure}{S\arabic{figure}}

\begin{center}
\text{\Large Supplemental Material}
\end{center}

\section{Eigenvalues and eigenvectors of a tridiagonal symmetric centrosymmetric matrix}
\begin{figure}
    \centering
    \includegraphics[width=.5\linewidth]{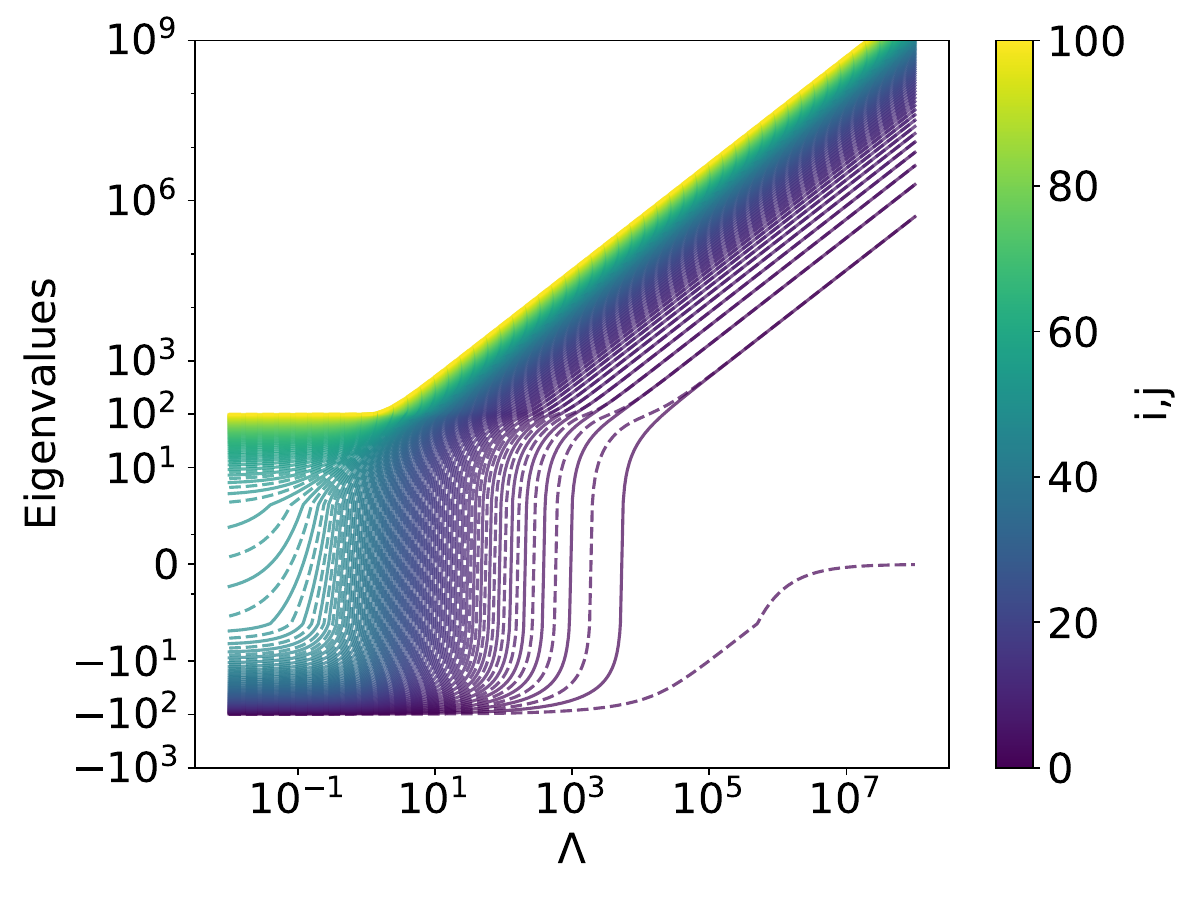}
    \caption{ Eigenvalues as a function of the dimensionless interaction strength $\Lambda=UN/J$ for $N=200$ particles. The eigenvectors, ordered in ascending order, are presented by bold lines ($\lambda_i$) and dashed lines ($\ell_j)$.}
    \label{fig:1}
\end{figure}

We are interested in solving the eigenvalue problem for the following tridiagonal symmetric centrosymmetric $(N+1)\times(N+1)$ matrix, with $N$ even
\begin{equation}
    \hat{H}=\begin{pmatrix}
\begin{array}{c|c|c}
\ A\ & \ \mathbf{x} &0\\
\hline
\ \mathbf{x}^T & \ 0 &\mathbf{x}^T\mathbb{J} \\
\hline
0 & \mathbb{J}\mathbf{x} & \mathbb{J}A\mathbb{J} \\
\end{array}
\end{pmatrix}
\end{equation}
where $A$ is a tridiagonal symmetric $\frac{N}{2}\times\frac{N}{2}$ matrix  characterized by $N-1$ real parameters, 
$\mathbf{x}$ is a 1-parameter $\frac{N}{2}$ -column vector 
\begin{equation}
    A=\begin{pmatrix}
        \begin{array}{ccccc}
        a_0&b_1&\\
        b_1&a_1&b_2& \\
        &\ddots&\ddots&\ddots\\
        &&b_{\frac{N}{2}-2}&a_{\frac{N}{2}-2}&b_{\frac{N}{2}-1} \\
        &&&b_{\frac{N}{2}-1}&a_{\frac{N}{2}-1} \\
        \end{array}
    \end{pmatrix}\quad \mathbf{x}=\begin{pmatrix}
        \begin{array}{c}
        0\\
        \vdots\\
        \vdots\\
        0\\
        b_{\frac{N}{2}}
        \end{array}
    \end{pmatrix}
\end{equation}
while $\mathbb{J}=\delta_{i,\frac{N}{2}-j}$ is the $\frac{N}{2}\times\frac{N}{2}$ exchange matrix. It has been proven \cite{Andrew1973,Cantoni1976}, that the eigenvalues of a tridiagonal symmetric matrix are distinct if the two secondary diagonals have not any zero entry. Furthermore, arranging them in descending order, they will be alternatively symmetric and skew-symmetric, starting with symmetric, as shown in Fig.~\ref{fig:1}. In particular, the skew-symmetric eigenvalues $\lambda_i$, with $i=1,\dots,\frac{N}{2}$, satisfy the same eigenvalue problem of the matrix $A$, i.e. 
\begin{equation}
 A\mathbf{u}_i=\lambda_i\mathbf{u}_i   
 \label{SSEP}
\end{equation}
 with $\{\mathbf{u}_i\}_{i=1,\dots,\frac{N}{2}}$ orthonormal set, and the corresponding eigenvectors of $\hat{H}$ are given by
\begin{equation}
    \mathbf{v}_i=\frac{1}{\sqrt{2}}\begin{pmatrix}
        \ \mathbf{u}_i\\
        0\\
        -\mathbb{J}\mathbf{u}_i
    \end{pmatrix}
\end{equation}
Vice versa, the symmetric eigenvalues $\ell_j$, with $j=0,\dots,\frac{N}{2}$, are given by the eigenvalues of
\begin{equation}
    \begin{pmatrix}
\begin{array}{c|c}
\ A\ & \ \sqrt{2}\mathbf{x} \\
\hline
 \sqrt{2}\mathbf{x}^T &  0
\end{array}
\end{pmatrix} 
\begin{pmatrix}
\begin{array}{c}
\ \mathbf{y}_j \\
 \alpha_j
\end{array}
\end{pmatrix}=\ell_j
\begin{pmatrix}
\begin{array}{c}
\ \mathbf{y}_j \\
 \alpha_j
\end{array}
\end{pmatrix}
\label{SEP}
\end{equation}
where $(\mathbf{y}_j\ \alpha_j)^T$ form an orthonormal set. Then, the eigenvectors of $\ell_j$ are given by
\begin{equation}
    \mathbf{w}_j=\frac{1}{\sqrt{2}}\begin{pmatrix}
        \ \mathbf{y}_j\\
        2\alpha_j\\
        \ \mathbb{J}\mathbf{y}_j
    \end{pmatrix}
\end{equation}
Moreover, the set $\{\mathbf{v}_1,\dots,\mathbf{v}_{\frac{N}{2}},\mathbf{w}_0,\mathbf{w}_1,\dots \mathbf{w}_{\frac{N}{2}}\}$ forms an orthonormal set. The matrices in Eq.$(\ref{SSEP})$ and Eq.$(\ref{SEP})$ are two tridiagonal symmetric matrices, their determinants can be computed from a three-term recurrence relation in terms of the continuant \cite{Hager1988}, in particular, the $n-$continuant is defined as
\begin{equation}
    f_n=a_{n-1}f_{n-1}-b^2_{n-1}f_{n-2}
\end{equation}
where $f_{-1}\equiv0$, $f_{0}\equiv1$,  $f_{1}\equiv a_0$. The eigenvalues of a $n\times n$ tridiagonal symmetric matrix are the zeroes of an $n$-th order polynomial related to the $n$-continuant. In particular, 
the eigenvalues of Eq.$(\ref{SSEP})$ are given by the polynomial equation
\begin{equation}
    p_\frac{N}{2}(\lambda_i)\equiv(\lambda_i-a_{\frac{N}{2}-1})p_{\frac{N}{2}-1}(\lambda_i)-b^2_{\frac{N}{2}-1}p_{\frac{N}{2}-2}(\lambda_i)=0
\end{equation}
while for the ones of Eq.$(\ref{SEP})$ we have
\begin{equation}
    p_{\frac{N}{2}+1}(\ell_j)\equiv\ell_jp_{\frac{N}{2}}(\ell_j)-b^2_{\frac{N}{2}}p_{\frac{N}{2}-1}(\ell_j)=0.
\end{equation}

\section{Estimates for the critical interaction strength for quasi-MQST}
\begin{figure}
    \includegraphics[width=.43\linewidth]{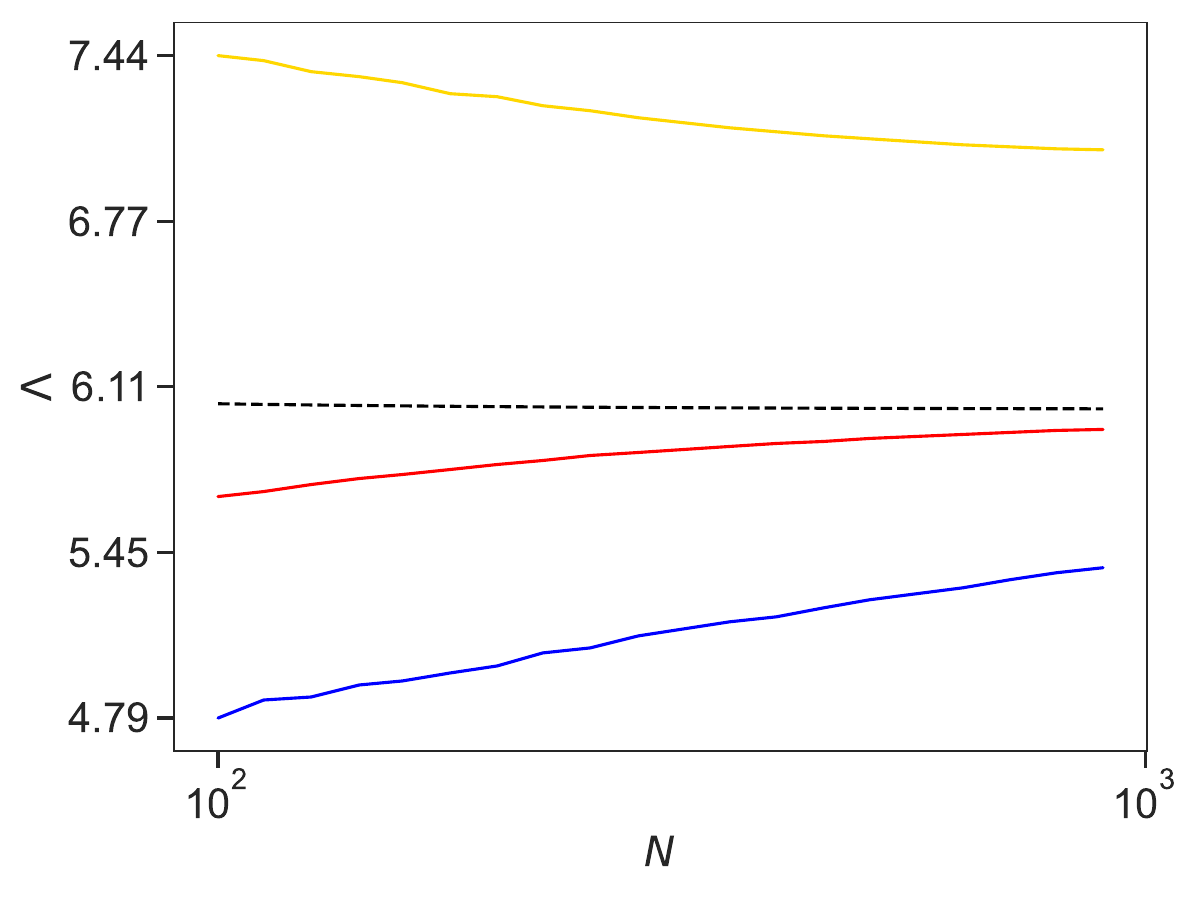}
    \includegraphics[width=.43\linewidth]{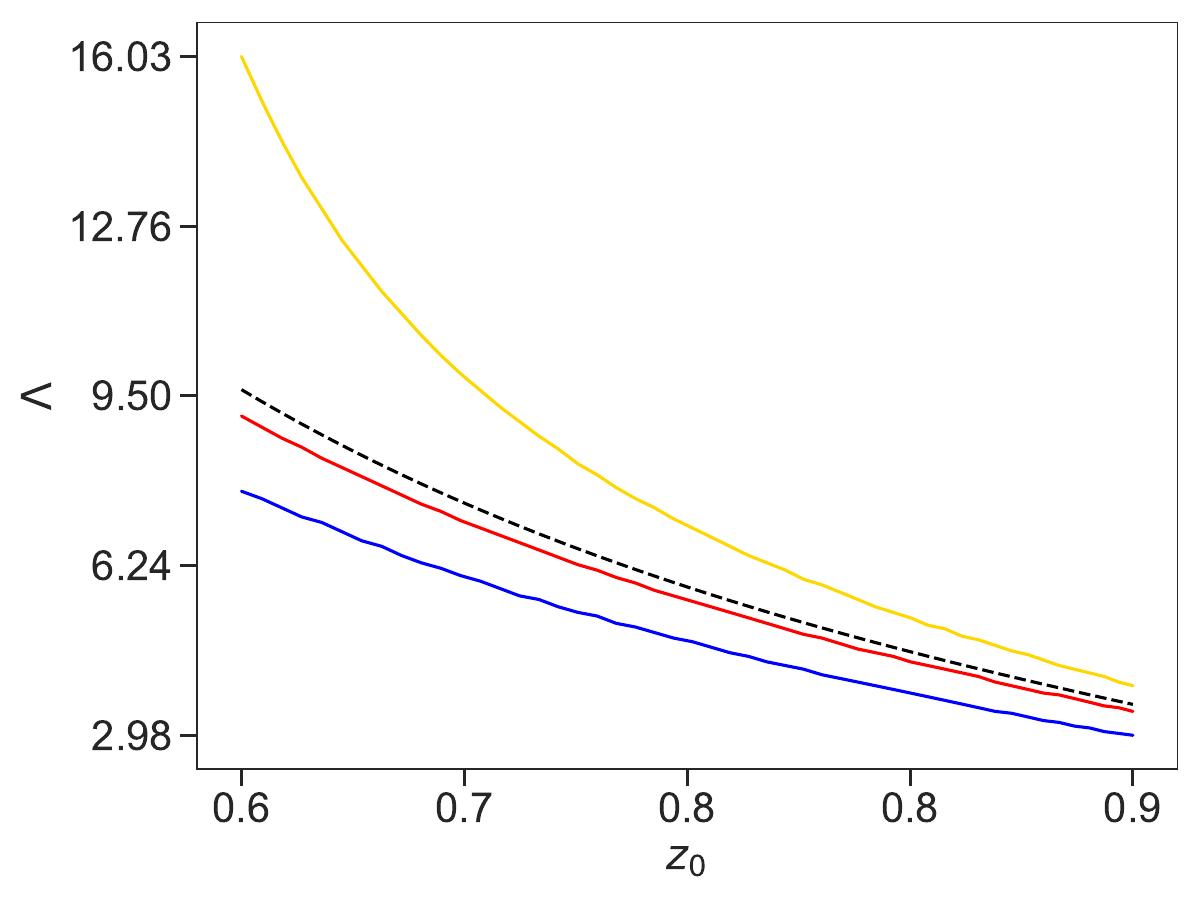}
    \caption{ Various estimates for the critical value for quasi-MQST, defined in the text, ($\Lambda_1$, blue;  $\bar{\Lambda}$, red; $\Lambda_2$, yellow; and $\Lambda_{c,\mathrm{MF}}$, black)  as a function of particle number $N$ at fixed $\tilde{z}_0 = 2 n_0/N=0.8$  (left panel)  and of initial population imbalance $z_0=\langle \hat z(0)\rangle $ at fixed particle number $N=150$ (right panel).  In both panels, the initial state of the quantum problem is the ground state of the Hamiltonian with $\Lambda_0 = 20$. 
}
    \label{fig:2}
\end{figure}

We propose here various estimates for the critical interaction strength for quasi-MQST, corresponding to the generalization of the mean-field critical value $\Lambda_{c,\mathrm{MF}}$. 
In the main text, we have identified the interaction strength $\bar{\Lambda}$ as the interaction strength where the function $\sigma^*(\Lambda)$ exhibits a jump discontinuity. Two additional quantum analogues of the mean-field estimate can be defined based on the cumulative weight of the contributions with $\sigma > \sigma_c$, namely $\sum_{\sigma>\sigma_c}A_{0,\sigma}C_{0,\sigma}$. The first, denoted $\Lambda_1$, is the smallest interaction strength at which this sum becomes nonzero, corresponding to the emergence of oscillations around a nonzero population imbalance. The second, $\Lambda_2$, is defined as the interaction strength at which these terms account for more than half of the total contribution. By construction, these parameters satisfy $\Lambda_1 \leq \bar{\Lambda} \leq \Lambda_2$, and in the mean-field limit ($N \gg 1$), they all converge to $\Lambda_{c,\mathrm{MF}}$, as shown in Fig.~\ref{fig:2} (left panel). In Fig.~\ref{fig:2} right panel we show their dependence  on the initial population imbalance $z_0$.  

\section{Quantum dynamics with atomic coherent states as initial state}

\begin{figure}
    \centering
    \includegraphics[width=.5\linewidth]{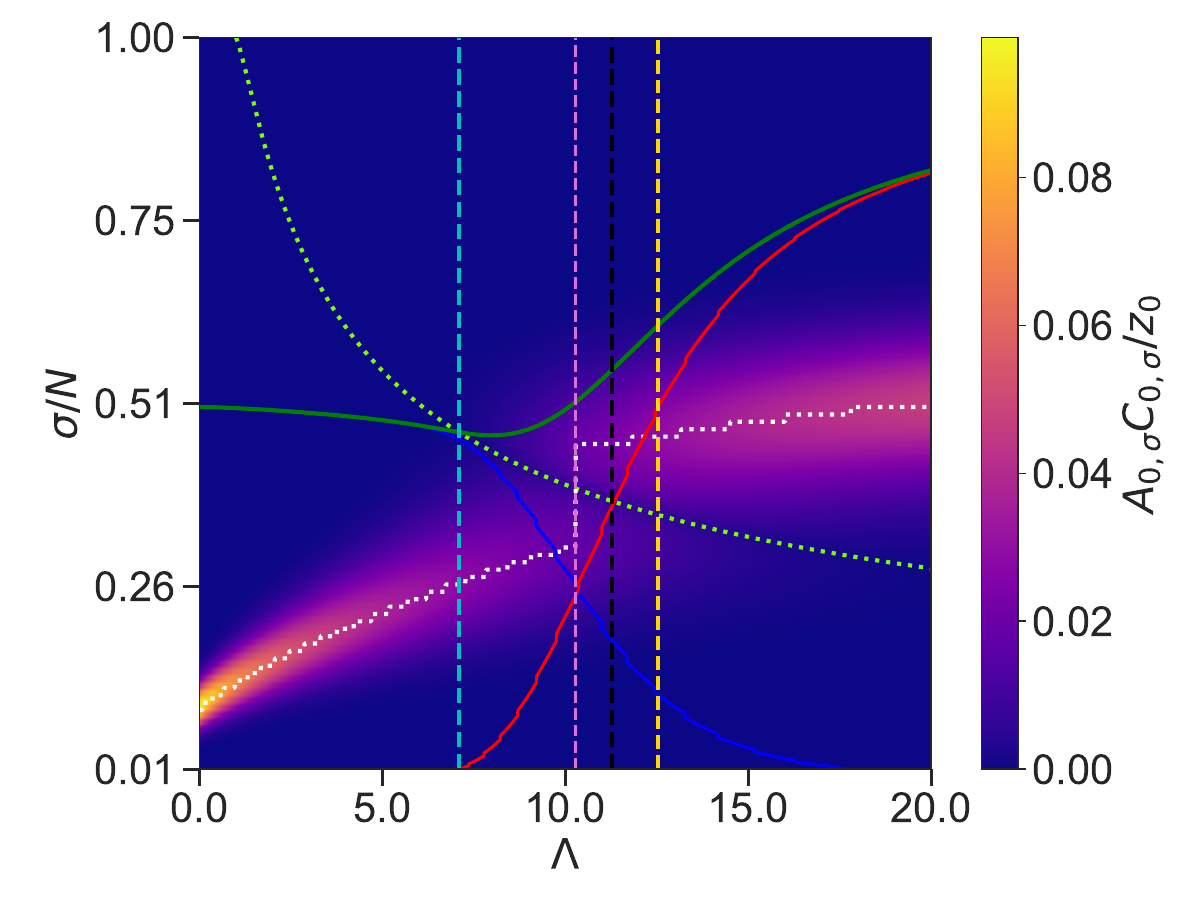}
    \caption{
Color map of the product of the coefficients $A_{0,\sigma}C_{0,\sigma}$, defined in Eqs.(9) and (10) of the main text, normalized by the initial population imbalance $z_0$, as a function of the dimensionless interaction strength $\Lambda$ and the normalized sum index $\sigma/N$ for a system with $N=200$ particles. The white dotted line indicates the value $\sigma^*/N$ corresponding to its maximum for each value of $\Lambda$.
The lime dotted line shows the critical sum index (see Eq.(13) of the main text). 
The three solid lines correspond to the sum of $A_{0,\sigma}C_{0,\sigma}/z_0$ over all $\sigma$ (blue), over $\sigma > \sigma_c$ (red), and over $\sigma < \sigma_c$ (green). 
The vertical dashed lines indicate   critical interaction strengths for MQST in mean field ($\Lambda_{c,\mathrm{MF}}$, black) and for quasi-MQST  ($\bar \Lambda$, magenta; $\Lambda_1$, cyan;  $\Lambda_2$ yellow).  The initial state is the atomic coherent state prepared with $z_0=0.57$ and $\phi_0=0$.
}
    \label{fig:4}
\end{figure}

\begin{figure}
    \centering
    \includegraphics[width=.5\linewidth]{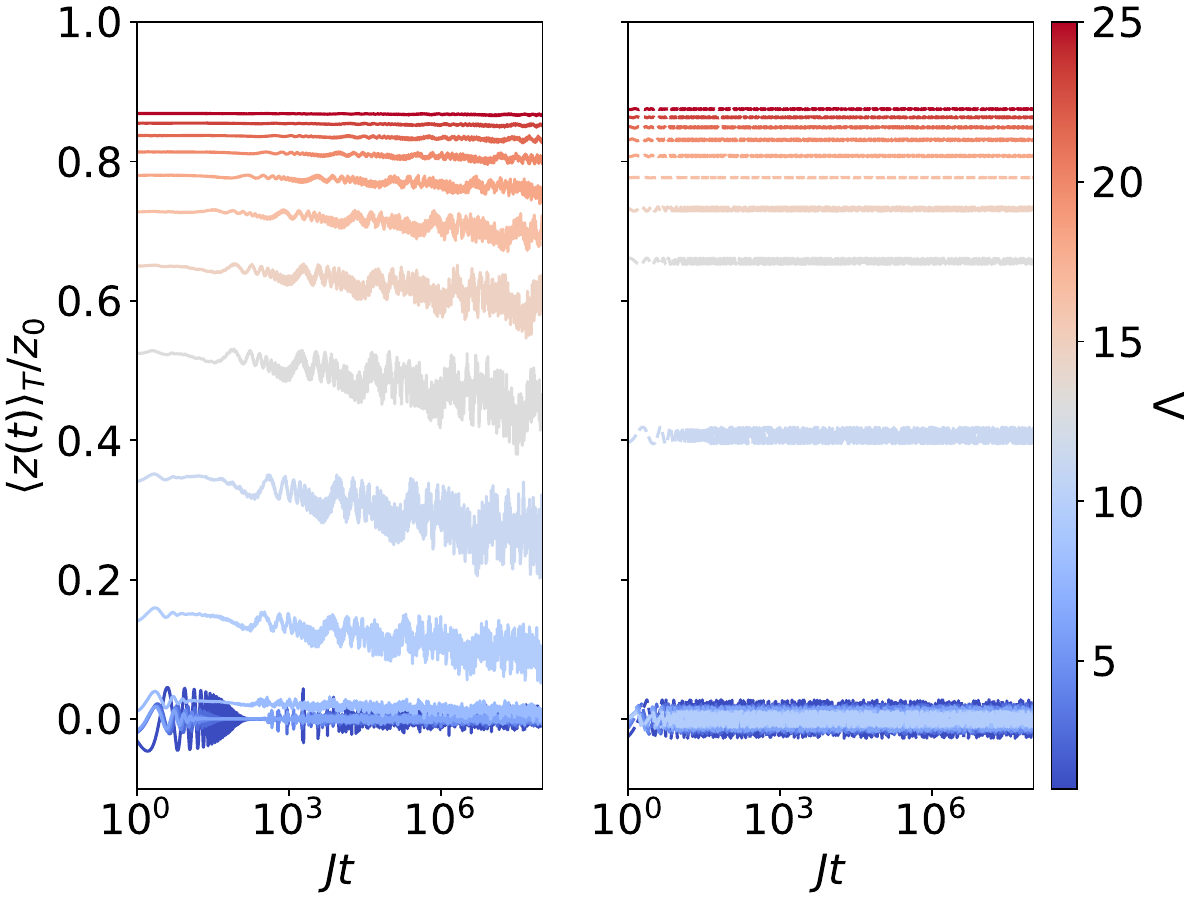}
    \caption{ Fully quantum  (left panel) and mean-field (right panel) time evolution of the time-averaged population imbalance   $\langle z(t)\rangle_T / z_0$ (dimensionless) as a function of time, in unites of $J^{-1}$, for a system of $N=200$ particles and initial imbalance $z_0 = 0.57$,  at various values of the interaction strength $\Lambda$  as indicated in the color-bar label.  The running averaging time  window is set to $JT = 25$. The initial state is the atomic coherent state prepared with $z(0)=z_0$ and $\phi_0=0$, and we use the same initial values  for the mean-field solution. }
    \label{fig:5}
\end{figure}

For completeness, we show here in Figs.~\ref{fig:4} and~\ref{fig:5} the counterparts of Figures 4 and 5 of the main text, obtained here by taking as initial state an atomic coherent state \cite{Arecchi1972,Gilmore1990}, Eq.(5) of the main text. We find that quasi-MQST also emerges in this case, showing that our conclusions are independent from the details of the initial state. 

\end{document}